\documentstyle[12pt]{article}
\def\be{\begin{equation}}
\def\ee{\end{equation}}
\def\bea{\begin{eqnarray}}
\def\eea{\end{eqnarray}}

\begin{document}

\begin{center}
{\Large{\bf Interaction of Moving Branes with Background Massless
and Tachyon Fields in Superstring Theory}}

\vskip .5cm {\large Zahra Rezaei and Davoud Kamani} 
\vskip .1cm {\it
Physics Department, Amirkabir University of Technology
(Tehran Polytechnic)\\
P.O.Box: 15875-4413, Tehran, Iran}\\
{\sl e-mails: z.rezaei , kamani@aut.ac.ir}\\
\end{center}

\begin{abstract}

Using the boundary state formalism we study a moving
D$p$-brane in a partially compact spacetime in the presence
of the background fields: Kalb-Ramond $B_{\mu\nu}$, a
$U(1)$ gauge field $A_{\alpha}$ and the tachyon field.
The boundary state enables us to obtain interaction
amplitude of two branes with above background fields.
The branes are parallel or perpendicular to each other.
Presence of the background fields, compactification
of some directions of the spacetime, motion of the
branes and arbitrariness of the branes' dimensions
give a general feature to the system. Due to the tachyon
fields and velocities of the branes, the behavior of the
interaction amplitude reveals obvious differences from
what is conventional.

\end{abstract}



\vskip .5cm
\newpage

\section{Introduction}

The discovery of the D-branes, as an inevitable part
of the string theory \cite{1}, induced to study the
properties and interactions of the branes. One of the
most applicable methods for this purpose, is the
boundary state formalism. Boundary state is a BRST
invariant state which describes the creation of closed
string from vacuum.

Among achievements in this formalism is extending it to the
superstring theory and considering the contribution of the conformal
and super conformal ghosts to the boundary state \cite{2}. There are
separate studies which add background fields such as the Kalb-Ramond
field $B_{\mu\nu}$, $U(1)$ gauge field in the compact spacetime
\cite{3} and the tachyon field \cite{4, 5, 6} to the subject of
boundary state. These background fields give a more general feature
to the subject. Apart from the longitudinal fluctuations of the
brane (for instance the $U(1)$ gauge field and tachyon field),
transverse fluctuations of the brane \cite{7} should also be
considered. This enables us to interpret it as a dynamical object.
This can be performed by considering velocity for the brane \cite{8,
9}. These motivated us to take into account all background fields
and also compactification of some directions of the spacetime to
study moving branes in a general framework of superstring theory.
This general set-up can not be found in the literatures of the
boundary state and branes' interaction.

Since open strings are quantum excitations of brane \cite{10},
presence of the open string tachyon reveals the instability of the
brane. In the bosonic string theory this is a natural property,
while in the superstring theories this occurs in special cases. For
instance there are D$p$-branes with wrong dimensions in the type IIA
and type IIB superstring theories; That is, there are D$p$-branes
with odd dimensions in the type IIA theory and even dimensions in
the type IIB theory! \cite{11} which are unstable. Actually this
instability can be removed by rolling of the tachyon toward its
minimum potential \cite{12}. During this process tachyon energy
dissipates to the bulk modes and an unstable system reach to a
stable state which consists of lower dimensional branes or just the
closed string vacuum without any D-brane \cite{10}. Usually in the
literature the tachyon field has been considered in just one
dimension and its effects have been studied on a space-filling
brane, while in the present paper we consider a D$p$-brane with
arbitrary dimension, and hence the tachyon field possesses
components along all directions of the brane worldvolume.

In this manuscript we calculate the boundary state
corresponding to a moving D$p$-brane in the presence
of the background fields $B_{\mu\nu}$, $U(1)$ gauge
field and tachyon. Then we use this boundary state to
detect the interaction between two moving D-branes.
There is no restriction on the branes' dimensions and
they can be parallel or perpendicular to each other.
To keep the generality, we let some
of the spacetime directions be compact. We will
observe that presence of the tachyon prevents the
closed string from wrapping around the compact directions.
Using the boundary state, we calculate the interaction
amplitude between two branes in the NS-NS and the R-R
sectors. Due to the presence of the velocities and
the background tachyon fields there is no cancellation
between these amplitudes. This occurs
even for the similar and parallel D$p$-branes with the
same background fields. We shall observe that the
interaction amplitude vanishes after a long time (or
equivalently for large distances of branes). The origin of
this effect is the rolling of the background tachyon
field and decaying of the D-branes in this limit.

So putting all these together allows us to study
a system in the most general feature to obtain
considerable results in spite of some mathematical
difficulties because of considering longitudinal
and transverse fluctuations simultaneously.
\section{The boundary state associated with a D$p$-brane}

To obtain the boundary state corresponding to a
moving brane in the presence of the antisymmetric
field $B_{\mu\nu}$ in bulk, and tachyon and $U(1)$
gauge fields on the boundary we consider the following
sigma-model action for closed string
\begin{eqnarray}
S =&~& -\frac{1}{4\pi\alpha'} {\int}_\Sigma
d^{2}\sigma(\sqrt{-g}g^{ab}G_{\mu\nu}\partial_a
X^{\mu}\partial_b X^{\nu}+\varepsilon^{ab}
B_{\mu\nu}\partial_a X^{\mu}\partial_b X^{\nu})
\nonumber\\
&~&
+\frac{1}{2\pi\alpha'} {\int}_{\partial\Sigma} d\sigma
\bigg{(} A_\alpha
\partial_{\sigma}X^{\alpha}+V^{i}X^{0}\partial_{\tau}X^{i}
+\frac{1}{2}U_{\alpha\beta}X^{\alpha}X^{\beta}\bigg{)},
\end{eqnarray}
where the first integral is on the worldsheet of the closed string,
exchanged between the branes, and the second one is on the boundary
of this worldsheet which can be at $\tau=0$ or $\tau=\tau_{0}$. The
$U(1)$ gauge field $A_{\alpha}$, lives on the D$p$-brane worldvolume
and $V^{i}$ is the brane velocity component along $X^{i}$-direction.
The set $\{X^{\alpha}\}$ and $\{X^{i}\}$ specify the directions
along and perpendicular to the D$p$-brane worldvolume, respectively.
The term $\frac{1}{2}U_{\alpha\beta}X^{\alpha}X^{\beta}$ with
constant symmetric matrix $U_{\alpha\beta}$ specifies the tachyon
profile. According to \cite{13} the tachyon field appears in a
square form in the action to produce a Gaussian integral. We take
the tachyon field to have components along the D$p$-brane
worldvolume. Here we consider $G_{\mu\nu}$ as the flat spacetime
metric with the signature $\eta_{\mu\nu}={\rm diag}
(-1,1,\cdot\cdot\cdot,1)$ and the Kalb-Ramond field $B_{\mu\nu}$ to
be a constant field.

Vanishing the variation of the action (1) with respect to
$X^{\mu}(\sigma,\tau)$, gives us the equations of motion as well as
the boundary equations of the emitted (absorbed) closed string.
\subsection{Bosonic part of the boundary state}

Boundary equations resulted from the action (1) at
$\tau=0$ are as in the following
\[
[\partial_{\tau} (X^{0}-V^{i}X^{i}) + {\cal {F}}^{0}_{\;\;\;
\alpha}\partial_{\sigma}X^{\alpha}-U^{0}_{\;\;\;\alpha}
X^{\alpha}]|B_{x},\tau=0\rangle =0,
\]
\[
(\partial_{\tau} X^{\bar{\alpha}} + {\cal
{F}}^{\bar{\alpha}}_{\;\;\;\beta}\partial_{\sigma}X^{\beta}
-U^{\bar{\alpha}}_{\;\;\;\beta}X^{\beta}) |B_{x}, \tau=0\rangle
=0,\]
\begin{equation}
(X^{i}-V^{i}X^{0}-y^{i})|B_{x}, \tau=0\rangle =0.
\end{equation}
In above boundary conditions $X^{\overline{\alpha}}$ shows the
spatial directions of the brane worldvolume (i.e.
$\overline{\alpha}\neq0$) and $\cal {F}$ is the total field
strength,
${\cal{F}}_{\alpha\beta}=\partial_{\alpha}A_{\beta}-\partial_{\beta}
A_{\alpha}-B_{\alpha\beta}$, which contains $B$ field as well as the
$U(1)$ gauge field. Note that we have assumed the mixed elements of the
Kalb-Ramond field to be zero, i.e, $B^{\alpha}_{\;\;\;i}=0$.

The solution of the closed string equation of motion is
\begin{equation}
X^{\mu}(\sigma,\tau)=x^{\mu}+2\alpha^{'}p^{\mu}\tau+
2L^{\mu}\sigma+\frac{i}{2}
\sqrt{2\alpha^{'}}\sum_{m\neq0}\frac{1}{m}(\alpha^{\mu}_{m}
e^{-2im(\tau-\sigma)}
+\tilde{\alpha}^{\mu}_{m}e^{-2im(\tau+\sigma)}).
\end{equation}
$L^{\mu}$ is zero for non-compact directions, and
$L^{\mu}=N^{\mu}R^{\mu}$ for the compact direction $X^\mu$ with
the compactification radius $R^{\mu}$ and the closed string winding
number $N^{\mu}$. The closed string center of mass momentum is
$p^{\mu}=\frac{M^{\mu}}{R^{\mu}}$, where $M^{\mu}$ is the momentum
number of it. Substituting this solution into the boundary equations
(2), gives them in terms of oscillators and zero modes.
During this process an interesting condition on the closed string
winding is obtained
\[
U^{\alpha}_{\;\;\;\overline{\beta}}L_{\rm op}^{\overline{\beta}}
|B_{x}, \tau=0 \rangle=0 .
\]
We assumed that there is no compactification along time
direction and and hence $L^0=0$.
In the case of invertiblity of the matrix
$U^{\overline{\alpha}}_{\;\;\;\overline{\beta}}$, this
equation reduces to
$L_{\rm op}^{\overline{\alpha}}|B_{x}, \tau=0 \rangle=0$.
Therefore, the presence of
the background tachyon field prevents closed string from wrapping
around compact directions which are parallel to the
brane worldvolume.

Utilizing the coherent state method \cite{14} to solve the boundary
equations (2) for oscillating modes leads to the following state
\begin{equation}
|B_{\rm osc},\tau=0\rangle=\prod^{\infty}_{n=1}[\det
M_{(n)}]^{-1}\exp\bigg {[}-\sum_{m=1}^{\infty}\bigg
{(}\frac{1}{m}\alpha^{\mu}_{-m}{\cal{S}}_{(m)\mu\nu}
\widetilde{\alpha}^{\nu}_{-m}\bigg {)}\bigg{]}|0\rangle,
\end{equation}
where the matrix ${\cal{S}}_{(m)}$ is defined by
\begin{eqnarray}
&~&
{\cal{S}}_{(m)}=S_{(m)}+{(S_{(-m)}^{-1})}^{T},
\nonumber\\
&~& S_{(m)}=M_{(m)}^{-1}N_{(m)}.
\end{eqnarray}
The matrices
$M_{(m)}$ and $N_{(m)}$ which are functions of background fields
are defined by
\begin{equation}
M^{\mu}_{(m)\;\nu}=\Omega^{\mu}_{\;\;\;\nu}-\frac{i}{2m}
U^{\alpha}_{\;\;\;\beta}\delta^{\mu}_{\;\;\;\alpha}
\delta^{\beta}_{\;\;\;\nu}
\end{equation}
where
\begin{eqnarray}
\left\{
\begin{array}{rcl}
&~&\Omega^{0}_{\;\;\;\mu}=
\delta^{0}_{\;\;\mu}-V^{i}{\delta^{i}}_{ \mu}-{{\cal
{F}}^{0}}_{\alpha}{\delta^{\alpha}}_{\mu}, \\
&~&\Omega^{\bar{\alpha}}_{\;\;\;\mu}=\delta^{\bar{\alpha}}_{\;\;\;\mu}
-{{\cal{F}}^{\bar{\alpha}}}_{\beta}{\delta^{\beta}}_{
\mu}, \\
&~& \Omega^{i}_{\;\;\;\mu}=\delta^{i}_{\;\;\;\mu}
-V^{i}{\delta^{0}}_{\mu},
\end{array}\right.
\end{eqnarray}
and
\begin{eqnarray}
\left\{ \begin{array}{rcl}
&~& N^{0}_{(m)\mu}={\delta^{0}}_{ \mu}-V^{i}{\delta^{i}}_{ \mu}+
{{\cal {F}}^{0}}_{\alpha}{\delta^{\alpha}}_{\mu}
+\frac{i}{2m}U^{0}_{\;\;\alpha}\delta^{\alpha}_{\;\;\mu},\\
&~& N^{\bar{\alpha}}_{(m)\mu}={\delta^{\bar{\alpha}}}_{\mu}
+{{\cal {F}}^{\bar{\alpha}}}_{\beta}{\delta^{\beta}}_{\mu}
+\frac{i}{2m}U^{\bar{\alpha}}_{\;\;\beta}
\delta^{\beta}_{\;\;\mu}, \\
&~& N^{i}_{(m)\mu}=-{\delta^{i}}_{ \mu}+V^{i}{\delta^{0}}_{ \mu}.
\end{array}\right.
\end{eqnarray}
When we solve the boundary equations, the matrix
${(S^{-1}_{(-m)})}^{T}$ also appears in the Eq. (5). This is due to
the fact that the matrix $S_{(m)}$ is mode-dependent and generally
is not orthogonal. In the absence of the tachyon field, $S$ becomes
mode-independent and orthogonal, so ${\cal{S}}=S$ \cite{3}. The
infinite product in the Eq. (4), which comes from path integral, can
be regularized \cite{15} as
\begin{equation}
\prod^{\infty}_{n=1}[\det M_{(n)}]^{-1}=\sqrt{\det
\Omega}\;\det\bigg{[}\Gamma \bigg{(}\frac{U}{1+2i\Omega}
\bigg{)}\bigg{]}.
\end{equation}

From now on we consider a selected direction $X^{i_{0}}$ for the
motion of the D$p$-brane and hence the other components of the
velocity are zero. We also define $V^{i_{0}}=V$. By this assumption,
the zero mode part of the boundary state becomes
\begin{eqnarray}
{|B_{x},\tau=0\rangle}^{(0)}=&~&\frac{T_{p}}{2}
{\int^{\infty}_{-\infty}}
\prod_{\alpha}dp^{\alpha}\;\bigg{\{} \exp\bigg{[}-4i\alpha'
(U^{-1})_{\alpha\beta}\bigg{(}(1-\frac{1}{2}
\delta_{\alpha\beta})
p^{\alpha}p^{\beta}+Vp^{i_{0}}p^{\beta}\delta^{\alpha}_{\;0}
\bigg{)}\bigg{]}
\nonumber\\
&~&\times\delta(x^{i_{0}}-Vx^{0}-y^{i_{0}}) \prod_{i'\neq
i_{0}}\delta(x^{i'}-y^{i'})
\nonumber\\
&~&\times\prod_{\alpha}|p^{\alpha}_{L} =p^{\alpha}_{\rm R}\rangle
\prod_{i'\neq
i_{0}}|p^{i'}_{L}=p^{i'}_{\rm R}=0\rangle
|p^{i_{0}}_{L}=p^{i_{0}}_{\rm R}=\frac{1}{2}Vp^{0}\rangle
\bigg{\}}.
\end{eqnarray}
Two delta functions indicate the position of the
brane along the perpendicular directions. The integration
over the momenta indicates that the effects of all values of
the momentum components have been taken into account.
In addition, the equality $p_L^\alpha = p_R^\alpha$ originates
from the un-wrapping of the closed string around the brane
directions and non-compactness of the time direction.

There are two special limiting cases for $U$. In the limit
$U_{\alpha\beta}\rightarrow0$, the oscillating part of the boundary
state, i.e. the Eq. (4), reduces to a boundary state corresponding
to a moving D$p$-brane in the absence of tachyon field, \cite{9}.

When we send some of the elements of $U$ to infinity, somehow
we are looking at the
boundary state in the concept of tachyon condensation. This
condensation can be performed on some or all elements of the
tachyon matrix $U$. Without loss of generality consider $U$ as a
diagonal matrix. By sending to infinity a spatial element
$U_{\overline{\alpha}\overline{\alpha}}$, the boundary state
transforms to the one related to a moving D$(p-1)$-brane which has
lost its dimension along the $X^{\bar \alpha}$-direction, and is
in the presence of a new tachyon field $U'_{(p-1)\times(p-1)}$
which does not include the component
$U_{\overline{\alpha}\overline{\alpha}}$.

Notable point here is that although in the process of condensation
along the $X^{\overline{\alpha}}$-direction the matrices $M$ and
$\cal{S}$ in boundary state (4) change to lower dimensional ones, as
expected, the effect of condensated component remains as a
$\sqrt{U_{\overline{\alpha}\overline{\alpha}}}$ factor after
regularization of infinite product $\prod^{\infty}_{n=1}[\det
M_{(n)}]^{-1}$. This result is different from the conventional case
in which this factor is canceled by the factor
$\frac{1}{\sqrt{U_{\overline{\alpha}\overline{\alpha}}}}$ from zero
mode part, that is absent here.

When condensation occurs along the time component of the tachyon
matrix, $U_{00}\rightarrow\infty$, beside the decreasing of the
brane worldvolume dimension in the $X^{0}$-direction, the brane loses its
velocity, too. In other words, the tachyon condensation
along the temporal direction fixes the D$p$-brane in time and
space, i.e. makes an instantonic D$p$-brane which has no
velocity.

\subsection{Fermionic part of the boundary state}

To find boundary equations for the fermionic degrees of freedom
there are two ways: 1) supersymmetrizing the action (1) and putting
the variation of the fermionic part of the action equal to zero; 2)
Since the supersymmetrized action is invariant under the global
worldsheet supersymmetry transformations, we can perform the
worldsheet supersymmetry on the bosonic boundary Eqs. (2) and
transform them to fermionic ones. Here we choose the second
approach. So the fermionic boundary equations are
\begin{eqnarray}
&~& [-i\eta(\psi^{0}_{+}-V^{i}\psi^{i}_{+})
+(\psi^{0}_{-}-V^{i}\psi^{i}_{-})+{\cal{F}}^{0}_{\;\;\;\alpha}
(-i\eta\psi^{\alpha}_{+}-\psi^{\alpha}_{-})
\nonumber\\
&~& -U^{0}_{\;\;\;\nu}(-i\eta\psi^{\nu}_{+}+\psi^{\nu}_{-})]|B_{\psi},
\eta, \tau=0\rangle=0,
\nonumber
\end{eqnarray}
\begin{eqnarray}
[-i\eta\psi^{\overline{\alpha}}_{+}
-\psi^{\overline{\alpha}}_{-}
+{\cal{F}}^{\overline{\alpha}}_{\;\;\;\beta}
(-i\eta\psi^{\beta}_{+}+\psi^{\beta}_{-})
-U^{\overline{\alpha}}_{\;\;\;\nu}(-i\eta\psi^{\nu}_{+}-
\psi^{\nu}_{-})]|B_{\psi}, \eta, \tau=0\rangle=0,
\nonumber
\end{eqnarray}
\begin{eqnarray}
[-i\eta (\psi^{i}_{+}-V^{i}\psi^{0}_{+})
- (\psi^{i}_{-}-V^{i}\psi^{0}_{-})]|B_{\psi},
\eta,\tau=0\rangle=0.
\end{eqnarray}
With respect to the solution of the equations of motion for the
fermions,
\begin{equation}
\psi^{\mu}_{-}=\sum_{k}
\psi^{\mu}_{k}e^{-2ik(\tau-\sigma)}\;\;\;\;,\;\;\;\;
\psi^{\mu}_{+}=\sum_{k}
\widetilde{\psi}^{\mu}_{k}e^{-2ik(\tau+\sigma)},
\end{equation}
the boundary state Eqs. (11) can be represented as
\begin{equation}
(\psi^{\mu}_{k}-i\eta
{S^{\mu}}_{(k)\nu}\tilde{\psi}^{\nu}_{-k})|B_{\psi},\eta,
\tau=0\rangle =0.
\end{equation}
Note that in the Eqs. (12) and (13), $k$ is an integer number $m$
for the R-R sector,
$\psi^{\mu}_{m}=d^{\mu}_{m}$ and
$\widetilde{\psi}^{\mu}_{m}=\widetilde{d}^{\mu}_{m}$
while in the NS-NS sector, $k$ is a half-integer number $r$ with
$\psi^{\mu}_{r}=b^{\mu}_{r}$ and
$\widetilde{\psi}^{\mu}_{r}=\widetilde{b}^{\mu}_{r}$. The constant
number $\eta$ can be $+1$ or $-1$. No matter we choose $+1$ or $-1$,
because for gaining the interaction of the branes, we need to use
the boundary state which has been affected by the GSO projector. As
will be seen, this projection operator causes the both states with
$\eta=+1$ and $\eta=-1$ to contribute to the interaction.

Similar to the bosonic part we should also consider the portion of
the super conformal ghosts in the fermionic boundary state. The
super ghosts include the commuting fields $\beta$, $\gamma$,
$\widetilde{\beta}$ and $\widetilde{\gamma}$.
\subsubsection{The NS-NS Sector}

According to the Eq. (13), the resultant NS-NS sector boundary state of
the fermions is given by
\begin{equation}
|B_{\psi},\eta,\tau=0\rangle_{\rm NS}=\prod^{\infty}_{r=1/2}[\det
M_{(r)}]\exp\bigg {[}i\eta\sum_{r=1/2}^{\infty}
(b^{\mu}_{-r}{\cal{S}}_{(r)\mu\nu}\widetilde{b}^{\nu}_{-r})
\bigg{]}|0\rangle_{\rm NS}.
\end{equation}
When the path integral is computed the determinant will be reversed
in comparing to the bosonic case, the Eq. (4).
This is due to the Grassmann nature of
the integration variables \cite{2}. As $r$ is half integer,
regularization of this infinite product is
\begin{equation}
\prod^{\infty}_{r=1/2}[\det M_{(r)}]=\det\bigg{(}\frac{\sqrt{\pi}}{
\Gamma[\frac{U}{2i\Omega}+\frac{1}{2}]}\bigg{)}.
\end{equation}

\subsubsection{The R-R Sector}

For acquiring the boundary state in the
R-R sector, we have to follow the same procedure of the NS-NS sector
with a bit difference which needs a careful notice. Since $k=m$ in
the Eq. (13) runs over integers in the R-R sector, there is a zero mode
which affects the boundary state. Solving the Eq. (13)
in the R-R sector, yields the following boundary state
\begin{equation}
|B_{\psi},\eta,\tau=0\rangle_{\rm R}=\prod^{\infty}_{m=1}[\det
M_{(m)}]\exp\bigg {[}i\eta\sum_{m=1}^{\infty}
(d^{\mu}_{-m}{\cal{S}}_{(m)\mu\nu}\widetilde{d}
^{\nu}_{-m})\bigg{]}|B_{\psi},\eta\rangle^{(0)}_{\rm R}.
\end{equation}
Since $m$ is an integer number, regularization of the infinite
product is exactly similar to the bosonic case (of course here
the determinant is inverse of the bosonic case)
\begin{equation}
\prod^{\infty}_{m=1}[\det M_{(m)}]=\bigg{\{}\sqrt{\det
\Omega}\;\det\bigg{(}\Gamma \bigg{[} 1+\frac{U}{2i\Omega}
\bigg{]}\bigg{)}\bigg{\}}^{-1}.
\end{equation}

The state $|B_{\psi},\eta \rangle^{(0)}_{\rm R}$ in the Eq. (16)
is the zero mode boundary state
\begin{equation}
|B_{\psi},\eta \rangle^{(0)}_{\rm
R}=\bigg{[}C\Gamma_{11}\bigg{(}\frac{1+i\eta\Gamma_{11}}
{1+i\eta}\bigg{)}\exp
(\frac{1}{2}\Phi_{\mu\nu}\Gamma^{\mu}\Gamma^{\nu})\bigg{]}^{AB}|A\rangle
|\widetilde{B}\rangle,
\end{equation}
where $|A\rangle |\widetilde{B}\rangle$ is the vacuum of the zero
modes $d^{\mu}_{0}$ and $\widetilde{d}^{\mu}_{0}$. $C$ is the
charge conjugate matrix, and the antisymmetric matrix $\Phi$ is
defined in terms of the matrix ${\cal {S}}$,
\begin{equation}
{\cal {S}}=(1-\Phi)^{-1}(1+\Phi).
\end{equation}
Details of obtaining the Eqs. (18) and (19) is shown in the appendix A. Since
the matrix ${\cal {S}}$ should be orthogonal ${\cal {S}}^{-1}={\cal
{S}}^{T}$, its definition
${\cal{S}}_{(m)}=S_{(m)}+[(S_{(-m)})^{-1}]^{T}$ implies that the
matrix $S$ should satisfy the following relation
\begin{equation}
S_{(m)}^{T}-S_{(m)}^{-1}=S_{(-m)}^{T}+S_{(-m)}^{-1}.
\end{equation}
According to the Eqs. (5)-(8) $S$ is defined in terms of the
background fields. Thus, the Eq. (20) imposes a relation between
these background fields. When in the action (1) the tachyon and
velocity are put to zero, we receive $\Phi=\cal{F}$ and hence the
term $\exp(\frac{1}{2}{\Phi}_{\alpha\beta}
\Gamma^{\alpha}\Gamma^{\beta})$ reduces to the known
$\exp(\frac{1}{2}{\cal{F}}_{\alpha\beta}
\Gamma^{\alpha}\Gamma^{\beta})$ \cite{3}.

\section{Interaction of the branes}

Interaction amplitude between D$p_1$ and D$p_2$-branes in each
sector is defined by ${\cal{A}}_{\rm
NS-NS,R-R}=2\alpha'\int^{\infty}_{0}dt\;_{\rm NS,R}\langle
B^1,\tau=0|e^{-tH_{\rm NS,R}}|B^{2},\tau=0\rangle_{\rm NS,R}$. Total
Hamiltonian $H_{\rm NS,R}$ is sum of the Hamiltonians of
$X^{\mu}$'s, $\psi^{\mu}$'s, ghosts and superghosts in each sector.
For calculation of the interaction amplitude we need the total
projected boundary state. The total boundary state of each sector is
\[|B,\eta,\tau=0\rangle_{\rm NS,R}=|B_{X},\tau=0\rangle|B_{\rm gh},
\tau=0\rangle|B_{\psi},\eta,\tau=0\rangle_{\rm NS,R}|B_{\rm sgh},
\eta,\tau=0\rangle_{\rm NS,R}.\] In the appendix B the projection
process is discussed. Thus, the total projected boundary states
find the feature of the Eqs. (40) and (41).

\subsection{Interaction amplitude in the NS-NS sector}

Using the boundary state (40) for NS-NS sector, after a long
calculation the total interaction amplitude in this sector is
acquired as
\begin{eqnarray}
{\cal {A}}_{\rm NS-NS} = &~&
\frac{\alpha'V_{\overline{u}}}{8(2\pi)^{d_{\overline{i}}}}\;
\frac{T_{p_1}T_{p_2}}{|V_1-V_2|}
\prod^{\infty}_{m=1}\frac{\det[
M_{(m-1/2)1}M_{(m-1/2)2}]}{\det[M_{(m)1}M_{(m)2}]}
\nonumber\\
&~&\times{\int_{0}}^{\infty}dt \;
\bigg{\{}\prod_{\bar{i}_{c}}\Theta_{3}\bigg{(}\frac{y_{1}
^{\bar{i}_{c}}-y_{2}^{\bar{i}_{c}}}{2\pi
R_{\bar{i}_{c}}}\mid\frac{i\alpha't}{\pi
(R_{\bar{i}_{c}})^{2}}\bigg{)}
\nonumber\\
&~& \times
\bigg{(}\sqrt{\frac{\pi}{\alpha't}}\bigg{)}^{d_{\overline{i}_{n}}}
\exp \bigg{(}-\frac{1}{4\alpha't}\sum_{\overline{i}_{n}}
({y_{1}}^{\overline{i}_{n}}-{y_{2}}^{\overline{i}_{n}})^{2}
\bigg{)}
\nonumber\\
&~& \times\frac{1}{q}\bigg{(}\prod^{\infty}_{m=1}\bigg{[}
{\bigg{(}\frac{1-q^{2m}}{1+q^{2m-1}}
\bigg{)}}^{2}\frac{\det(1+{\cal{S}}_{(m-1/2)1}
{\cal{S}}^{T}_{(m-1/2)2}q^{2m-1})}
{\det(1-{\cal{S}}_{(m)1}{\cal{S}}^{T}_{(m)2}q^{2m})}\bigg{]}
\nonumber\\
&~& -\prod^{\infty}_{m=1}\bigg{[}{\bigg{(}\frac{1-q^{2m}}
{1-q^{2m-1}}
\bigg{)}}^{2}\frac{\det(1-{\cal{S}}_{(m-1/2)1}
{\cal{S}}^{T}_{(m-1/2)2}q^{2m-1})}
{\det(1-{\cal{S}}_{(m)1}{\cal{S}}^{T}_{(m)2}q^{2m})}\bigg{]}
\bigg{)}\nonumber\\
&~& \times \frac{1}{\sqrt{\det Q \;\det K_{1}\;\det
K_{2}}}\nonumber\\
&~& \times\exp\bigg{[}-\frac{1}{4}\bigg{(}E^{T}Q^{-1}E+
\sum_{\alpha'_{1},\beta'_{1}}
y_{2}^{\alpha'_{1}}y_{2}^{\beta'_{1}}(K_{1}^{-1})_
{\alpha'_{1}\beta'_{1}}
+\sum_{\alpha'_{2},\beta'_{2}}
y_{1}^{\alpha'_{2}}y_{1}^{\beta'_{2}}(K_{2}^{-1})_
{\alpha'_{2}\beta'_{2}}\bigg{)}\bigg{]}\bigg{\}},
\end{eqnarray}
where $q=e^{-2t}$, and $V_{\bar u}$ is common volume of the branes.
The set $\{\overline{i}\}$ shows directions perpendicular to both
branes except $i_{0}$, $\{\overline{u}\}$ is for the directions
along both branes except $0$, $\{\alpha'_{1}\}$ is used for the
directions along the D$p_1$-brane and perpendicular to the
D$p_2$-brane and $\{\alpha'_{2}\}$ indicates the directions along
the D$p_2$-brane and perpendicular to the D$p_1$-brane.
$\overline{i}_{c}$ and $\overline{i}_{n}$ are related to the compact
and non-compact parts of $\overline{i}$, respectively. The matrices
$Q$, $K_{1}$, $K_{2}$ and the doublet $E$ are defined through their
elements as in the following
\begin{eqnarray}
\left\{
\begin{array}{rcl}
&~& Q_{11}=\frac{\alpha't}{2(V_{2}-V_{1})^{2}}(1+{V_{1}}^{2})
(1-{V_{2}}^{2})+2i\alpha'(U^{-1}_{1})^{00}(1-V^2_2)^2, \\
&~& Q_{22}=\frac{\alpha't}{2(V_{2}-V_{1})^{2}}(1+{V_{2}}^{2})
(1-{V_{1}}^{2})-2i\alpha'(U^{-1}_{2})^{00}(1-V_1^2)^2, \\
&~& Q_{12}=Q_{21}=\frac{\alpha't}{(V_{2}-V_{1})^{2}}(1+{V_{1}}^{2})
(1+{V_{2}}^{2})(1-V_{1}V_{2}),
\end{array}\right.
\end{eqnarray}
\begin{eqnarray}
\left\{ \begin{array}{rcl} &~&
E_{1}=\frac{i}{2(V_{2}-V_{1})}\bigg{[}{y_{2}}^{i_{0}}
(1+{V_{1}}^{2})^{2}
-{y_{1}}^{i_{0}}(1+V_{1}V_{2})\bigg{]},
\\ &~& E_{2}=\frac{i}{2(V_{2}-V_{1})}\bigg{[}{y_{1}}^{i_{0}}
(1+{V_{2}}^{2})^{2}-{y_{2}}^{i_{0}}(1+V_{1}V_{2})\bigg{]},
\end{array}\right.
\end{eqnarray}
\begin{eqnarray}
\left\{ \begin{array}{rcl}
&~& K^{\alpha'_{1}\beta'_{1}}_{1}
=4i\alpha'(1-\frac{1}{2}\delta_{\alpha'_{1}
\beta'_{1}})(U^{-1}_{1})^{\alpha'_{1}\beta'_{1}}
-\alpha't\delta^{\alpha'_{1}\beta'_{1}}, \\
&~&
K^{\overline{u}\overline{v}}_{1}
=4i\alpha'(1-\frac{1}{2}\delta_{\overline{u}\overline{v}})
(U^{-1}_{1})^{\overline{u}\overline{v}}
-\frac{1}{2}\alpha't\delta^{\overline{u}\overline{v}}, \\
&~& K^{\alpha'_{1}\overline{u}}_{1}=K^{\overline{u}\alpha'_{1}}_{1}=
4i\alpha'(U^{-1}_{1})^{\alpha'_{1}\overline{u}}.
\end{array}\right.
\end{eqnarray}
Under the exchanges $1\rightarrow2$ and $i\rightarrow-i$ in
the elements of the matrix $K_{1}$ we receive the elements
of the matrix $K_{2}$.  As is obvious $Q$, $E$ and $K$'s are
completely velocity and tachyon dependent.

In the amplitude (21), the theta function comes from the compact
part of the set $\{X^{\bar i}\}$, while the exponential and its
pre-factor in the third line originate from the non-compact part of
it. In fact, the exponential is a damping factor with respect to the
distance of the branes. If all directions $\{X^{\bar i}\}$ are
compact the exponential and its pre-factor disappear. In this case
${\bar i}_c$ takes all values of ${\bar i}$. In the case that all
directions $\{X^{\bar i}\}$ are non-compact the $\Theta_3$-factor is
removed, hence ${\bar i}_n$ takes all values of ${\bar i}$. The next
two lines which contain the $\cal{S}$ matrix reflect the portion of
the oscillators, conformal ghosts and super conformal ghosts. The
remaining part, which is obtained by integration over the momenta,
the Eq. (10), is due to the presence of the velocities and the
background tachyon fields. In absence of the velocities and tachyon
fields, this factor is absent too and hence the interaction
amplitude resembles to the one in \cite{3}.
\subsection{Interaction amplitude in the R-R sector}

For interaction amplitude in the R-R sector we use the total GSO
projected boundary state for the R-R sector, the Eq. (41), and
follow the same procedure in the NS-NS sector, so
\begin{eqnarray}
{\cal
{A}}_{\rm R-R} =&~&
\frac{\alpha'V_{\overline{u}}}{8(2\pi)^{d_{\overline{i}}}}\;
\frac{T_{p_1}T_{p_2}}{|V_1-V_2|}{\int_{0}}^{\infty}dt
\;
\bigg{\{}\prod_{\bar{i}_{c}}\Theta_{3}\bigg{(}\frac{y_{1}
^{\bar{i}_{c}}-y_{2}^{\bar{i}_{c}}}{2\pi
R_{\bar{i}_{c}}}\mid\frac{i\alpha't}{\pi
(R_{\bar{i}_{c}})^{2}}\bigg{)}
\nonumber\\
&~&\times \bigg{(}
\sqrt{\frac{\pi}{\alpha't}}\bigg{)}^{d_{\overline{i}_{n}}}
\exp \bigg{(}-\frac{1}{4\alpha't}\sum_{\overline{i}_{n}}
({y_{1}}^{\overline{i}_{n}}-{y_{2}}^{\overline{i}_{n}})^{2}
\bigg{)}
\nonumber\\
&~&\times\bigg{(}\zeta\prod^{\infty}_{m=1}\bigg{[}
{\bigg{(}\frac{1-q^{2m}}{1+q^{2m}}
\bigg{)}}^{2}\frac{\det(1+{\cal{S}}_{(m)1}
{\cal{S}}^{T}_{(m)2}q^{2m})}
{\det(1-{\cal{S}}_{(m)1}{\cal{S}}^{T}_{(m)2}q^{2m})}
\bigg{]}+\zeta'\bigg{)}
\nonumber\\
&~&\times \frac{1}{\sqrt{\det Q \;\det K_{1}\;\det
K_{2}}}
\nonumber\\
&~& \times\exp\bigg{[}-\frac{1}{4}\bigg{(}E^{T}Q^{-1}E+
\sum_{\alpha'_{1},\beta'_{1}}
y_{2}^{\alpha'_{1}}y_{2}^{\beta'_{1}}(K_{1}^{-1})_
{\alpha'_{1}\beta'_{1}}
+\sum_{\alpha'_{2},\beta'_{2}}
y_{1}^{\alpha'_{2}}y_{1}^{\beta'_{2}}(K_{2}^{-1})_
{\alpha'_{2}\beta'_{2}}\bigg{)}\bigg{]}\bigg{\}},
\end{eqnarray}
where
\begin{equation}
\zeta\equiv-\frac{1}{2} {\rm Tr}[G_{1}C^{-1}G^{T}_{2}C],
\end{equation}
\begin{equation}
\zeta'\equiv -i
{\rm Tr}[G_{1}C^{-1}G^{T}_{2}C\Gamma_{11}],
\end{equation}
and $G_{1,2}=\exp[\frac{1}{2}(\Phi_{(1,2)})_
{\mu\nu}\Gamma^{\mu}\Gamma^{\nu}]$. Note that the variables $\zeta$
and $\zeta'$ implicitly depend on the branes' dimensions through
$\Phi_1$ and $\Phi_2$ in $G_1$ and $G_2$.

Now we are eager to study the total amplitude, i.e. the
combination of the amplitudes in the NS-NS and R-R sectors. Consider
the following special case: there is no compactification, the two D$p$-branes
are parallel with the same dimensions, and the same fields living
on them. Thus, as in the literature, this interaction amplitude becomes zero
due to the cancellation of attractive and repulsive forces in the NS-NS
and R-R sectors, respectively.

In the case at hand, beside the living fields on the branes, the
velocities which are transverse fluctuations of the branes are
present, too. In the amplitude (21) and (25) the relative speed
appeared in the denominators. This puts a constraint on the system
that the velocities of the branes should be different, otherwise the
total amplitude becomes infinite. In this case, we can not check the
vanishing of the interaction amplitude for identical parallel branes
with the same fields. Therefore, even if all the fields are
identical, the velocities should be different. This causes the
branes to have different $\Phi$'s and consequently different
$\cal{S}$'s. So the NS-NS and R-R amplitudes cannot cancel the
effect of each other.
\section{Long distance behavior of the amplitude}

Now we find the interaction between the branes when they are far
from each other. That is, we find the behavior of the interaction
amplitudes (21) and (25) when time goes to infinity. Conventionally,
in the large distance only the massless states of the closed string
contribute to the branes interaction.

The large distance amplitude is equivalent to the long time behavior
of the branes. It can be acquired by sending $q$ to zero in the Eqs.
(21) and (25). So the interaction amplitudes due to massless states
in the NS-NS and R-R sectors are
\begin{eqnarray}
\lim_{q\rightarrow0}{\cal{A}}_{\rm NS-NS}= &~&
\frac{V_{\overline{u}}T_{p_1}T_{p_2}}{4(2\pi)^{d_{\overline{i}}}}
\;\frac{i(-1)^{(p_1+p_2)/2}\;2^{d_{\overline{u}}+1/2}}
{\alpha'^{(p_1+p_2)/2}(1+{V_1}^{2})(1+{V_2}^{2})}\;
\prod^{\infty}_{m=1}\frac{\det[
M_{(m-1/2)1}M_{(m-1/2)2}]}{\det[M_{(m)1}M_{(m)2}]}
\nonumber\\
&~& \times \int^\infty
dt\;\bigg{\{}\bigg{(}\sqrt{\frac{\pi}{\alpha't}}\bigg{)}^{d_{\overline{i}_{n}}}
\exp \bigg{(}-\frac{1}{4\alpha't}\sum_{\overline{i}_{n}}
({y_{1}}^{\overline{i}_{n}}-{y_{2}}^{\overline{i}_{n}})^{2}\bigg{)}
\nonumber\\
&~& \times\lim_{t\rightarrow\infty}
\frac{2[{\rm Tr}{({\cal{S}}_{(1)1}{\cal{S}}^{T}_{(1)2})-2]}}
{t^{1+(p_1+p_2)/2}}\bigg{\}},
\end{eqnarray}
 and
\begin{eqnarray}
\lim_{q\rightarrow0}{\cal{A}}_{\rm R-R}= &~&
\frac{V_{\overline{u}}\;T_{p_1}T_{p_2}}{8(2\pi)
^{d_{\overline{i}}}}\;
\frac{i(-1)^{(p_1+p_2)/2}\;2^{d_{\overline{u}}+1/2}}
{\alpha'^{(p_1+p_2)/2}(1+{V_1}^{2})(1+{V_2}^{2})}
\nonumber\\
&~& \times \int^\infty
dt\;\bigg{\{}\bigg{(}\sqrt{\frac{\pi}{\alpha't}}\bigg{)}^{d_{\overline{i}_{n}}}
\exp \bigg{(}-\frac{1}{4\alpha't}\sum_{\overline{i}_{n}}
({y_{1}}^{\overline{i}_{n}}-{y_{2}}^{\overline{i}_{n}})^{2}\bigg{)}
\nonumber\\
&~& \times\lim_{t\rightarrow\infty}
\frac{1}{t^{1+(p_1+p_2)/2}}\bigg{\}}.
\end{eqnarray}
We do not put the limit on the exponential part and its pre-factor
in the Eqs. (28) and (29) because these factors are related to the
positions of the branes, and closed string emission is independent
of the locations of the branes. When there is no tachyonic
background \cite{3}, last factors in the Eqs. (28) and (29) do not
have the factor $1/t^{1+(p_1+p_2)/2}$. Thus, due to the presence of
tachyon fields, the interaction amplitude decreases in time. In
fact, the statement that for large distances of the branes the
massless closed string states dominate in the interaction is valid
until there is no tachyon backgrounds in the system.

There is an interpretation for this unusual behavior. In fact, the
open string tachyon background causes an instability in the system.
Therefore, after long enough time, by rolling of the tachyon
\cite{12} toward its minimum potential, unstable D-branes decay to
the bulk modes and their dimensions decrease to reach a stable
system. Final products of this process are branes with lower
dimensions or vacuum of the closed string \cite{10}. The latter
implies that there are no physical perturbative open string states
around the minimum of the potential. This is due to the fact that
the open string states live only on the branes. Thus, in the concept
of interactive branes, by passing the time which leads to tachyon
rolling and decreasing of their dimensions, the branes'
configuration distorts and prevents them from the interaction.

The amplitude ${\cal{A}}_{\rm NS-NS}$ in the Eq. (28) depends on the
background fields through the factor ${\rm
Tr}({\cal{S}}_{(1)1}{\cal{S}}^{T}_{(1)2})$ and the determinants of
the matrices $\{M_{(m-1/2)}|m=1,2,3, \cdot\cdot\cdot\}$, while such
a dependence is absent in the amplitude ${\cal{A}}_{\rm R-R}$, the
Eq. (29). In other words, when the branes are far from each other,
the R-R amplitude becomes background independent.

Another interesting feature of the long time amplitude is
its time-dependent behavior on the branes' dimensions.
An exception here is the D-instanton. When two D-instantons,
which have the dimension $p_1=p_2=-1$, interact the
factor $1/t^{1+(p_1+p_2)/2}$ is removed and the amplitude
behavior in long time is resembled to a system without
tachyon. For this system the presence of the
tachyon does not affect the conventional behavior of the
large distance interaction.
\section{Conclusions}

The boundary state of a closed superstring traveling between two
moving branes in the presence of $B_{\mu\nu}$, tachyon and $U(1)$
gauge field was calculated. Notable feature in the boundary state
equations is the prevention of the closed string wrapping around the
compact directions of spacetime, which is due to the presence of the
tachyon field. As well, the boundary state includes a momentum
dependent exponential factor which is absent in the conventional
boundary states. This factor originates from the zero mode parts of
the velocity and tachyon terms in the boundary action.

The interaction amplitude of the branes via exchanging
of closed string was calculated for the NS-NS and R-R
sectors. It is shown that even for co-dimension parallel
branes with similar external fields, the total amplitude
is not zero. This is due to the presence of the velocities
and tachyon fields in the system.

The long distance behavior of the interaction amplitude was studied.
In this domain the instability of the branes, due to the background
tachyon fields, weakens the interaction. This decreasing behavior
can be understood by dissipation of the branes to the bulk modes
because of the rolling of the tachyon to its minimum potential in
long time regime. The interaction for two D-instantons obviates this
decreasing behavior. Thus, the long time amplitude in this case
behaves like the conventional case in which the massless states
dominate.
\begin{center}
\textbf{Appendix A}
\end{center}
\begin{center}
\textbf{Zero mode boundary state in the R-R sector}
\end{center}

The state $|B_{\psi},\eta \rangle^{(0)}_{\rm R}$ in the Eq. (16)
is the zero mode boundary state that obeys the following equation
\begin{equation}
|B_{\psi},\eta \rangle^{(0)}_{\rm R}={\cal{M}}^{(\eta)_{AB}}
|A\rangle |\widetilde{B}\rangle,
\end{equation}
where $|A\rangle |\widetilde{B}\rangle$ is the vacuum of the zero
modes $d^{\mu}_{0}$ and $\widetilde{d}^{\mu}_{0}$. The matrix
${\cal{M}}^{(\eta)}$ has to satisfy the equation
\begin{equation}
(\Gamma^{\mu})^{T}{\cal {M}}^{(\eta)}-i\eta
{\cal{S}}^{\mu}_{(m)\nu}\Gamma_{11}{\cal
{M}}^{(\eta)}\Gamma^{\nu}=0.
\end{equation}
Consider a solution in the form
\begin{equation}
{\cal {M}}^{(\eta)}=C\Gamma_{11}\bigg{(}\frac{1+i\eta\Gamma_{11}}
{1+i\eta}\bigg{)}G,
\end{equation}
in which $C$ is the charge conjugate matrix. Substitution of the Eq.
(32) into the Eq. (31) leads to the following equation for the
matrix $G$,
\begin{equation}
\Gamma^{\mu}G={\cal {S}}^{\mu}_{\;\;\;\nu}G\Gamma^{\nu}.
\end{equation}
There is a conventional solution for $G$ as
\begin{equation}
G= \exp (\frac{1}{2}\Phi_{\mu\nu}\Gamma^{\mu}\Gamma^{\nu}).
\end{equation}
Indeed one must expand the exponential with the convention that
all gamma matrices anticommute, therefore there are a finite
number of terms. The antisymmetric matrix $\Phi$ is defined in
terms of the matrix ${\cal {S}}$,
\begin{equation}
{\cal {S}}=(1-\Phi)^{-1}(1+\Phi).
\end{equation}

\begin{center}
\textbf{Appendix B}
\end{center}
\begin{center}
\textbf{GSO projected and ghosts boundary states}
\end{center}

The GSO projected boundary states are given by
\begin{equation}
|B,\tau=0\rangle_{\rm NS}=\frac{1-(-1)^{F+G}}{2}\frac{1-(-1)
^{\widetilde{F}+\widetilde{G}}}{2}|B,\eta=+1,\tau=0\rangle_{\rm
NS},
\end{equation}
\begin{equation}
|B,\tau=0\rangle_{\rm R}=\frac{1+(-1)^{n}(-1)^{F+G}}{2}
\frac{1-(-1)^{\widetilde{F}+\widetilde{G}}}{2}
|B,\eta=+1,\tau=0\rangle_{\rm R},
\end{equation}
where $n$ is an even
number for the type IIA superstring theory and is odd for the IIB
superstring theory. The definitions of $F$ and $G$,  are
\begin{equation}
F=\sum^\infty_{r=1/2} b^\mu_{-r} b_{r\mu}\;\;\;\;,\;\;\;\;
G=-\sum^\infty_{r=1/2}(\gamma_{-r}\beta_r+\beta_{-r}\gamma_r),
\end{equation}
for the NS-NS sector, and
\begin{equation}
(-1)^F=\Gamma_{11}(-1)^{\sum^\infty_{m=1} d^\mu_{-m}
d_{m\mu}}\;\;\;\;,\;\;\;\; G=-\gamma_0
\beta_0-\sum^\infty_{m=1}(\gamma_{-m}\beta_m+\beta_{-m}\gamma_m),
\end{equation}
for the R-R sector. Similar definitions also hold for ${\tilde F}$
and ${\tilde G}$. Thus, the total projected boundary states are
\begin{equation}
|B,\tau=0\rangle_{\rm NS}=\frac{1}{2}\bigg{(}|B,+,
\tau=0\rangle_{\rm NS}-|B,-,\tau=0\rangle_{\rm NS}\bigg{)},
\end{equation}
\begin{equation}
|B,\tau=0\rangle_{\rm R}=\frac{1}{2}\bigg{(}|B,+,
\tau=0\rangle_{\rm R}+|B,-,\tau=0\rangle_{\rm R}\bigg{)}.
\end{equation}

Since the bulk action in the Eq. (1) preserves conformal symmetry,
working in covariant formalism necessitates including conformal
ghosts \cite{2, 16}. In fact, what we need is the portion of ghosts
(i.e. anti-commuting fields $b$, $c$, $\widetilde{b}$ and
$\widetilde{c}$) in the bosonic boundary state. This part is
independent of the background fields and is expressed as
\begin{equation}
|B_{\rm gh},\tau=0\rangle=\exp\bigg{[}\sum^{\infty}_{m=1}
e^{4im\tau_{0}}
(c_{-m}\widetilde{b}_{-m}-b_{-m}\widetilde{c}_{-m})\bigg{]}
\frac{c_{0}
+\widetilde{c_{0}}}{2}|q=1\rangle|\widetilde{q}=1\rangle.
\end{equation}
In the superstring theory, in addition to the conformal ghosts,
we should also consider the super conformal ghosts. Thus, the
boundary state, corresponding to the super conformal
ghosts in the NS-NS and R-R sectors, are as in the following
\begin{equation}
|B_{\rm sgh},\eta,\tau=0\rangle_{\rm NS}=\exp\bigg{[}
i\eta\sum^{\infty}_{r=1/2}
(\gamma_{-r}\widetilde{\beta}_{-r}-\beta_{-r}
\widetilde{\gamma}_{-r})\bigg{]}
|P=-1\rangle|\widetilde{P}=-1\rangle.
\end{equation}
\begin{equation}
|B_{\rm sgh},\eta,\tau=0\rangle_{\rm R}=\exp\bigg{[}i\eta
\sum^{\infty}_{m=1}
(\gamma_{-m}\widetilde{\beta}_{-m}-\beta_{-m}
\widetilde{\gamma}_{-m})\bigg{]}|P=-\frac{1}{2}
\rangle|\widetilde{P}
=-\frac{3}{2}\rangle.
\end{equation}


\begin{thebibliography}{99}

\bibitem{1} J. Polchinski, Phys. Rev. Lett. 75 (1995) 4724.

\bibitem{2} C.G. Callan, C. Lovelace, C.R. Nappi and S.A.
Yost, Nucl. Phys. B288 (1987) 525;
Nucl. Phys. B293 (1987) 83; Nucl. Phys. B308 (1988) 221.

\bibitem{3} H. Arfaei and D. Kamani, Phys. Lett. B452
(1999) 54,  hep-th/9909167; Nucl. Phys. B561 (1999) 57-76,
hep-th/9911146.

\bibitem{4} T. Lee, Phys. Rev. D64 (2001) 106004; G. Arutyunov,
A. Pankiewicz and B. Stefanski jr, JHEP 06 (2001) 049; S.P. de
Alwis, Phys. Lett. B505 (2001) 215

\bibitem{5} E. T. Akhmedov, M. Laidlaw and
G. W. Semenoff, JETP Lett. 77: 1-6, 2003, PismaZh. Eksp. Teor. Fiz.
77: 3-8, 2003; M. Laidlaw, G. W. Semenoff, JHEP 0311:021, 2003; 11
M. Laidlaw, ``{\it On a Modification of the Boundary State Formalism
in Off-shell String Theory}", hep-th/0210270.

\bibitem{6} Z. Rezaei and D. Kamani, ``{\it Moving Branes 
in Presence of the Background Tachyon Fields}'', 
J. Exp. Theor. Phys. 140 (2011), arXiv: 1106.2097 [hep-th];
``{\it Moving Branes with Background Massless and Tachyon
Fields in the Compact Spacetime}'', arXiv: 1107.0380.

\bibitem{7} C.G. Callan and I.R. Klebanov, Nucl. Phys.
B465 (1996) 473.

\bibitem{8} M. Billo, P. Di Vecchia and D. Cangemi, Phys.
Lett. B400 (1997) 63.

\bibitem{9} D. Kamani, Mod. Phys. Lett. A15 (2000) 1655-1664,
hep-th/9910043.
\bibitem{10} A. Sen, Int. J. Mod. Phys. A20 (2005) 5513.

\bibitem{11} M. R. Gaberdiel, Class. Quant. Grav. 17 (2000) 3483-3520

\bibitem{12} A. Sen, JHEP 0204 (2002) 048.

\bibitem{13} D. Kutasov, M. Marino and G. Moore,
hep-th/0010108; JHEP 0010 (2000) 045.

\bibitem{14} M. Green, J. Schwarz and E. Witten,
``{\it Superstring theory}",
Vols. I and II (Cambridge University Press, 1987).

\bibitem{15} P. Kraus and F. Larsen, Phys. Rev. D63 (2001) 106004.

\bibitem{16} T. Uesugi, ``{\it Worldsheet Description of Tachyon
Condensation in Open String Theory}", hep-th/0302125.

\end{thebibliography}
\end{document}